\documentclass[newstyle,twocolumn,proceedings]{rmaa}
\renewcommand{\P}[1]{%
\ifnum#1=1\hbox{OW~168--326E}\fi
\ifnum#1=2\hbox{OW~167--317}\fi
\ifnum#1=3\hbox{OW~163--317}\fi
\ifnum#1=5\hbox{OW~158--323}\fi
\ifnum#1=0\hbox{OW~171--334}\fi}
\makeatletter
\title{Optical and near IR spectroscopy of low excitation HII regions with GTC}
\author{Enrique P\'erez-Montero, Marcelo Castellanos and \'Angeles I. D\'\i az 
\affil{Universidad Aut\'onoma de Madrid.}}

\fulladdresses{
\item  Departamento de F\'\i sica Te\'orica. C-XI. Universidad Aut\'onoma de Madrid. 28049. Cantoblanco. Madrid. Spain.(enrique.perez, marcelo.castellanos, angeles.diaz@uam.es).}

\shortauthor{P\'EREZ-MONTERO ET AL.}
\shorttitle{Low excitation HII regions spectroscopy}

\keywords{Galaxies:  Abundances --- HII regions}

\resumen{La determinaci\'on de abundancias qu\'\i micas en regiones de gas ionizado se basa en la detecci\'on de las l\'\i neas de emisi\'on de tipo auroral, a partir de las cuales se pueden deducir las temperaturas de l\'\i nea de distintos iones. Dichas l\'\i neas son d\'ebiles en objetos de baja excitaci\'on, por lo que su medici\'on en muchos de ellos es s\'olo posible en telescopios de gran abertura. El GTC permitir\'\i a llevar a cabo un estudio exhaustivo de abundancias qu\'\i micas en dichos objetos a partir de espectros de media-alta resoluci\'on, lo que acotar\'\i a a\'un m\'as las restricciones observacionales a los modelos de evoluci\'on qu\'\i mica en distintos tipos de galaxias.}

\abstract{The determination of chemical abundances in ionized gas nebulae is based on the detection of auroral emission lines, from which it is possible to deduce the line temperatures of several ions. These lines are weak in low excitation objects, so its measurement is only possible in great aperture telescopes. The GTC would allow to plan a comprehenssive study of chemical abundances in such objects from mid-high resolution spectra, what would give more accurate observational restrictions to models of chemical evolution. }

\begin{document}

\maketitle


The introduction and use of both a new generation of ground-based telescopes of 8-10 metres diameter and more sensitive detectors is going to allow the observations and subsequent analysis of a wide set of astrophysical objects with low surface brightness and/or at large distances that, until now, were not accesible. These new instruments can, as well, enhance the quality of the spectra and images of other objects that, althought observed, have not been completely studied due to the lack of data with the required signal-to-noise ratio.

The determination of chemical abundances in ionized gas nebulae is an example of what these instruments could improve. In the so-called direct method the chemical composition of the HII regions is deduced from the value of the electron temperature, that can be calculated from the measured intensity line ratio of certain ions (ex: [OIII]: $(4959\AA + 5007\AA)/4363\AA$, [SIII]: $(9069\AA + 9532\AA)/6312\AA$). Nevertheless, as the faint auroral lines are intrinsically weak they are only measurable in objects of high excitation and/or at not very large distances.

This can be seen in figure 1, where the spectra of both high (the upper panel) and low excitation (lower) HII regions are shown (Castellanos et al., 2002)

\begin{figure*}
   \begin{center}
    \includegraphics[width=4.1cm]{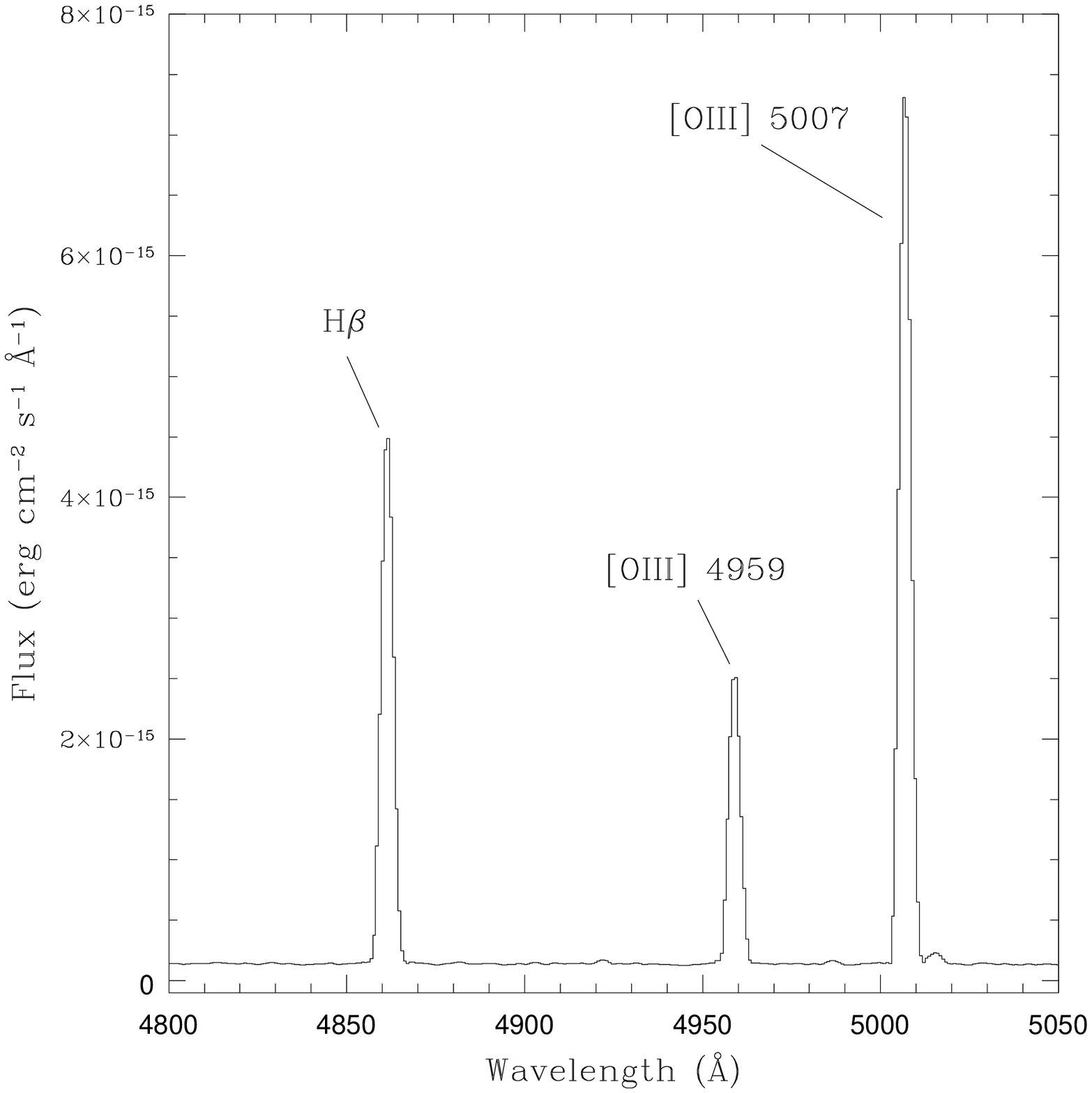}
    \includegraphics[width=4.1cm]{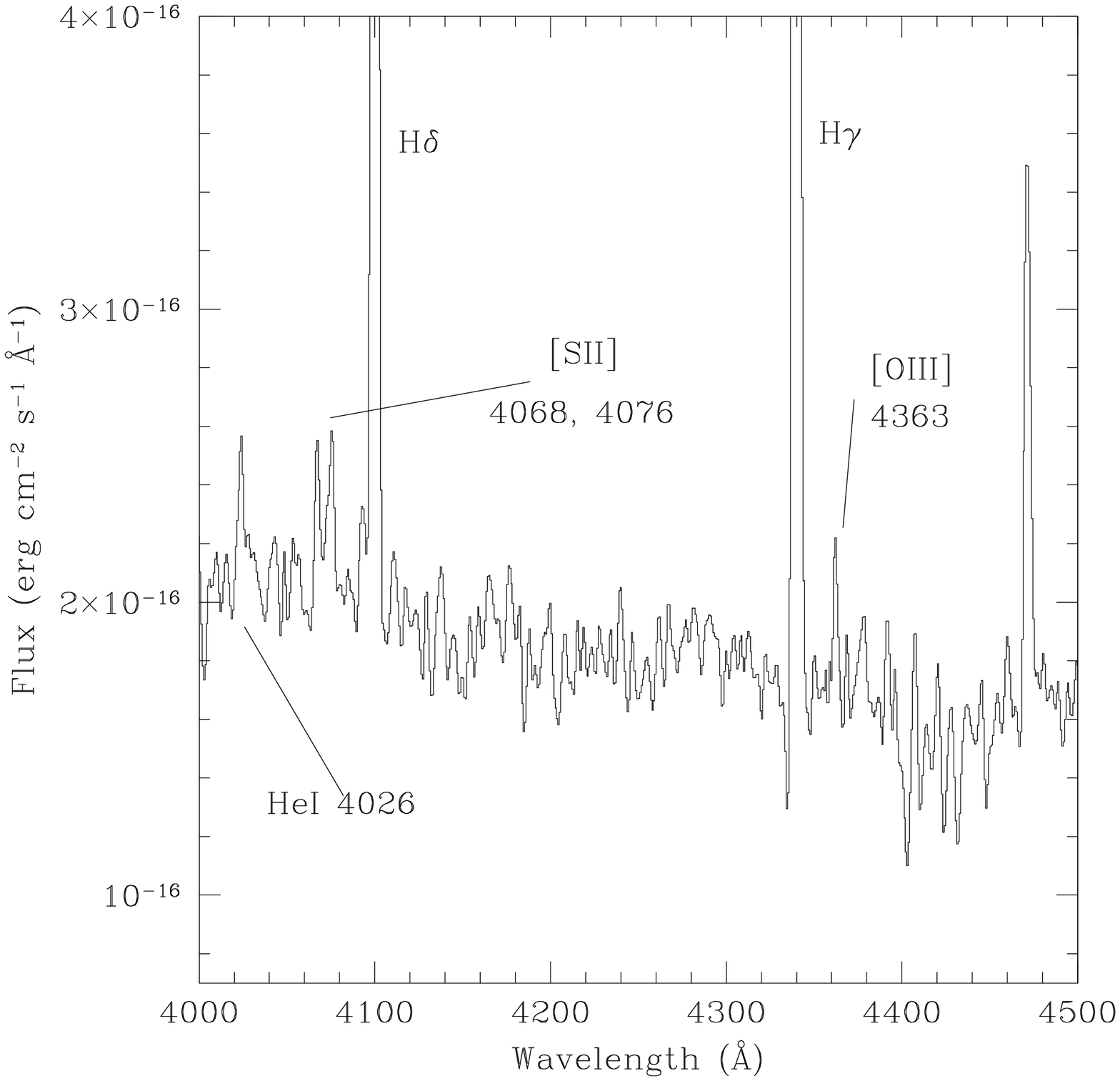}
    \includegraphics[width=4.1cm]{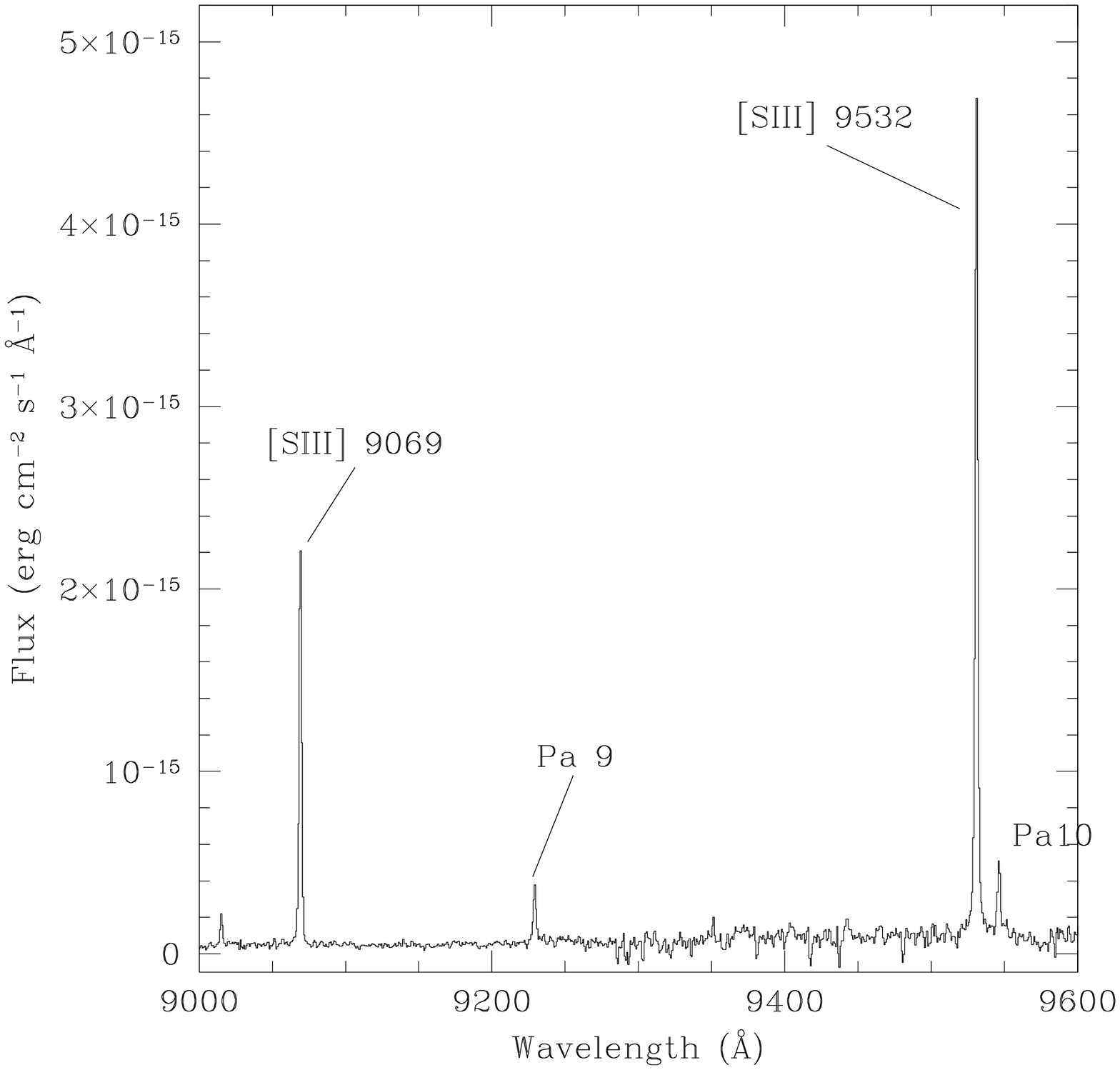}
    \includegraphics[width=4.1cm]{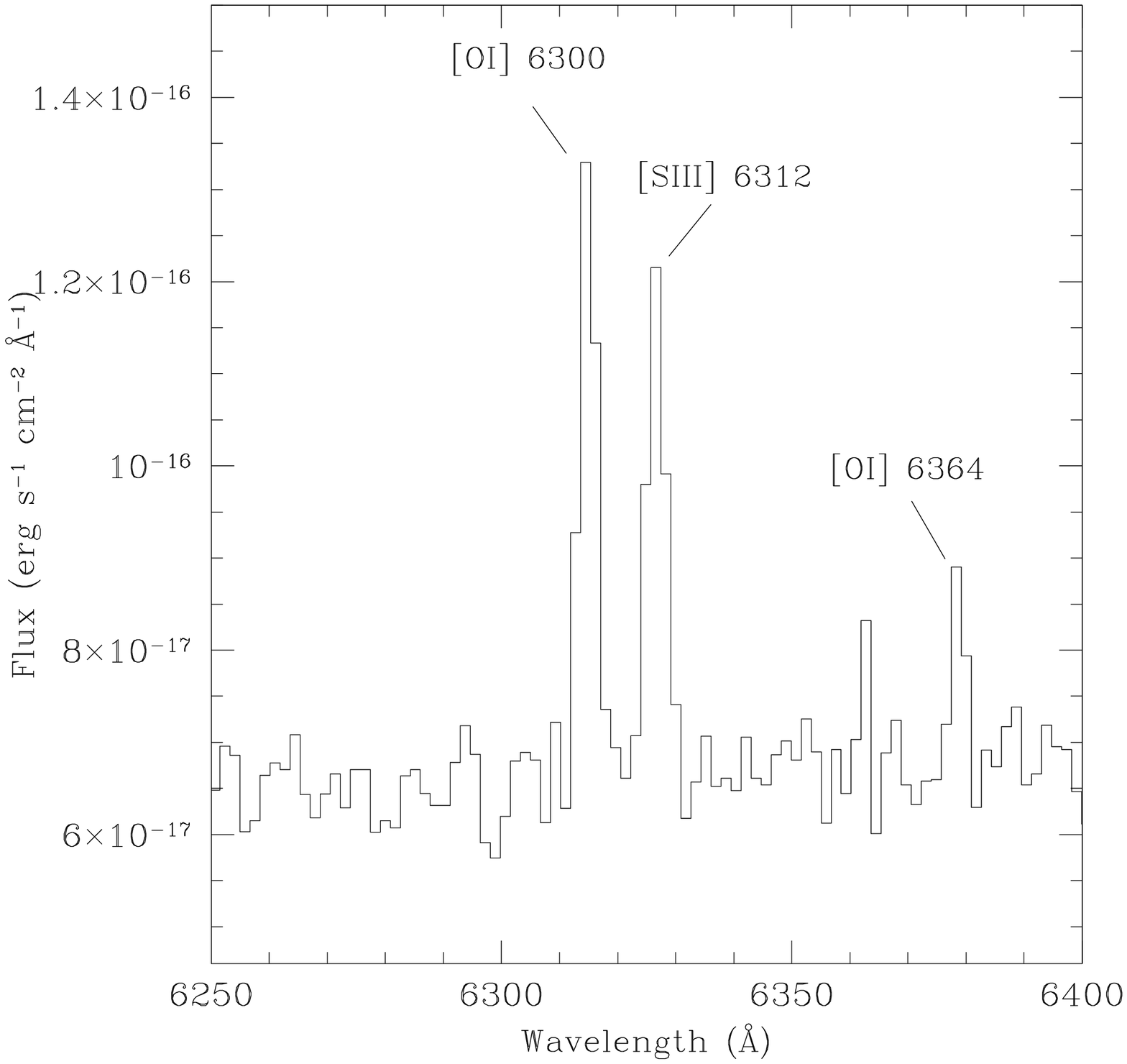}
   \end{center}
\vspace{-0.6cm}
  \begin{center}
\includegraphics[width=4.1cm]{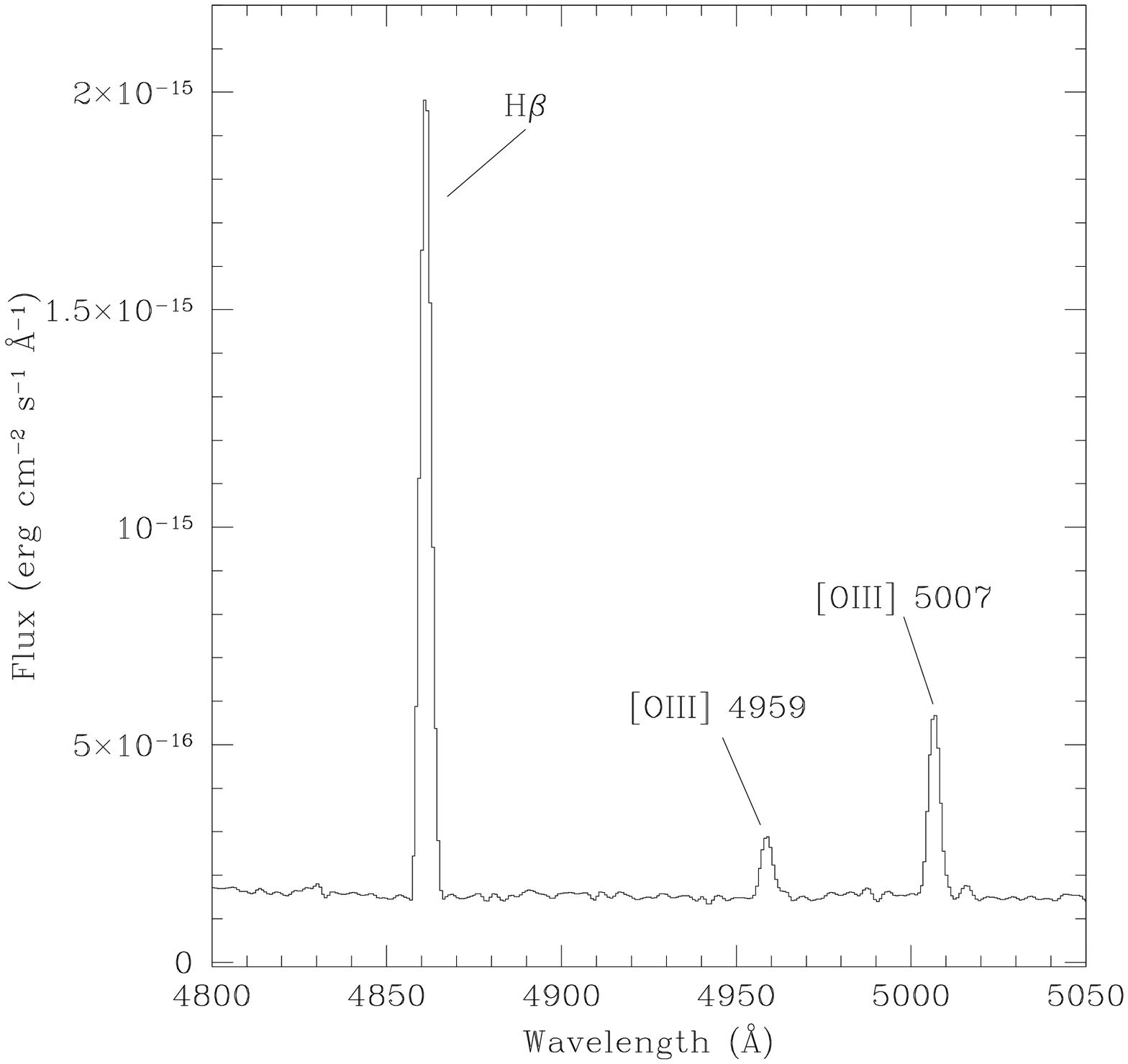}
\includegraphics[width=4.1cm]{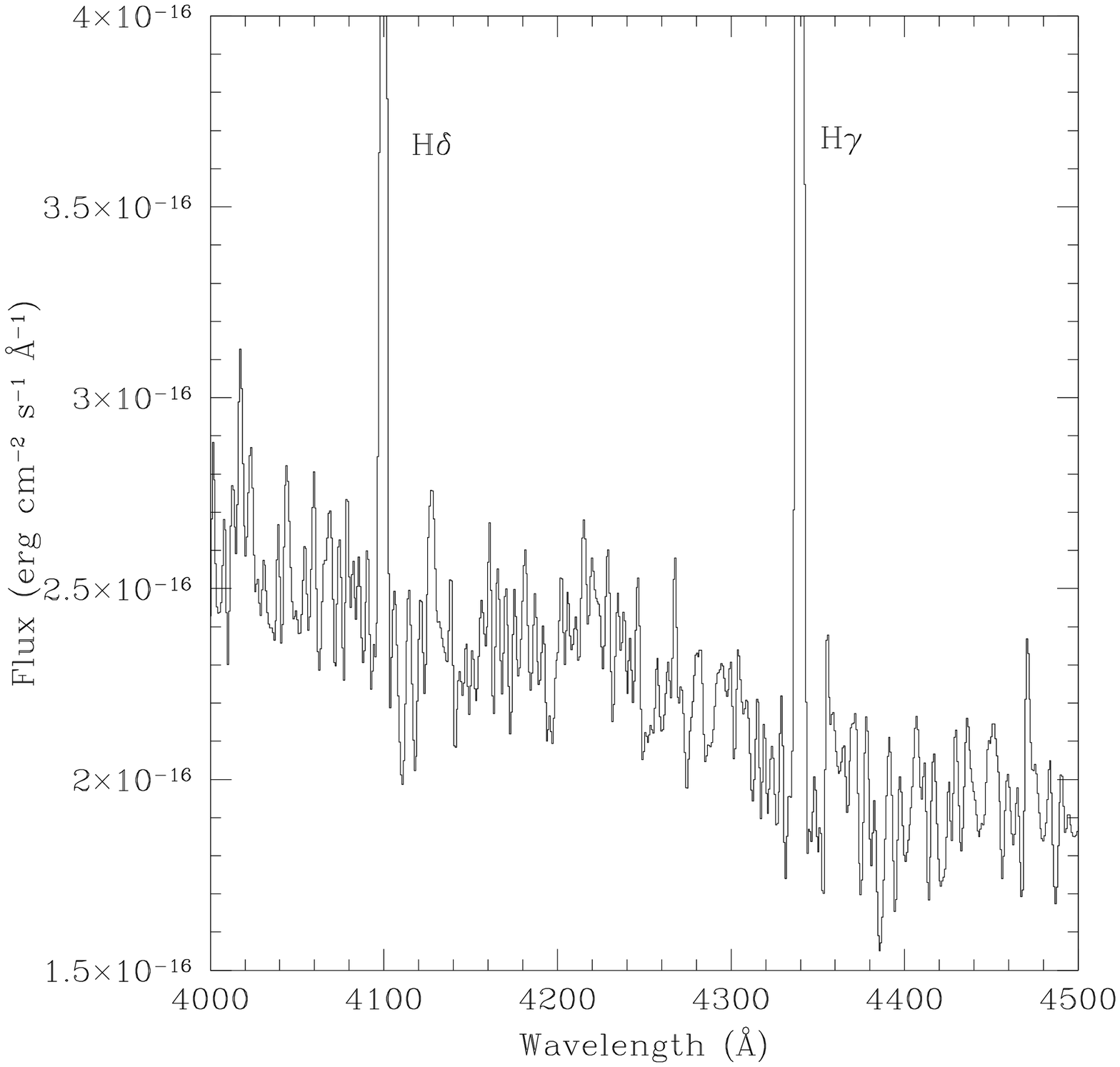}
\includegraphics[width=4.1cm]{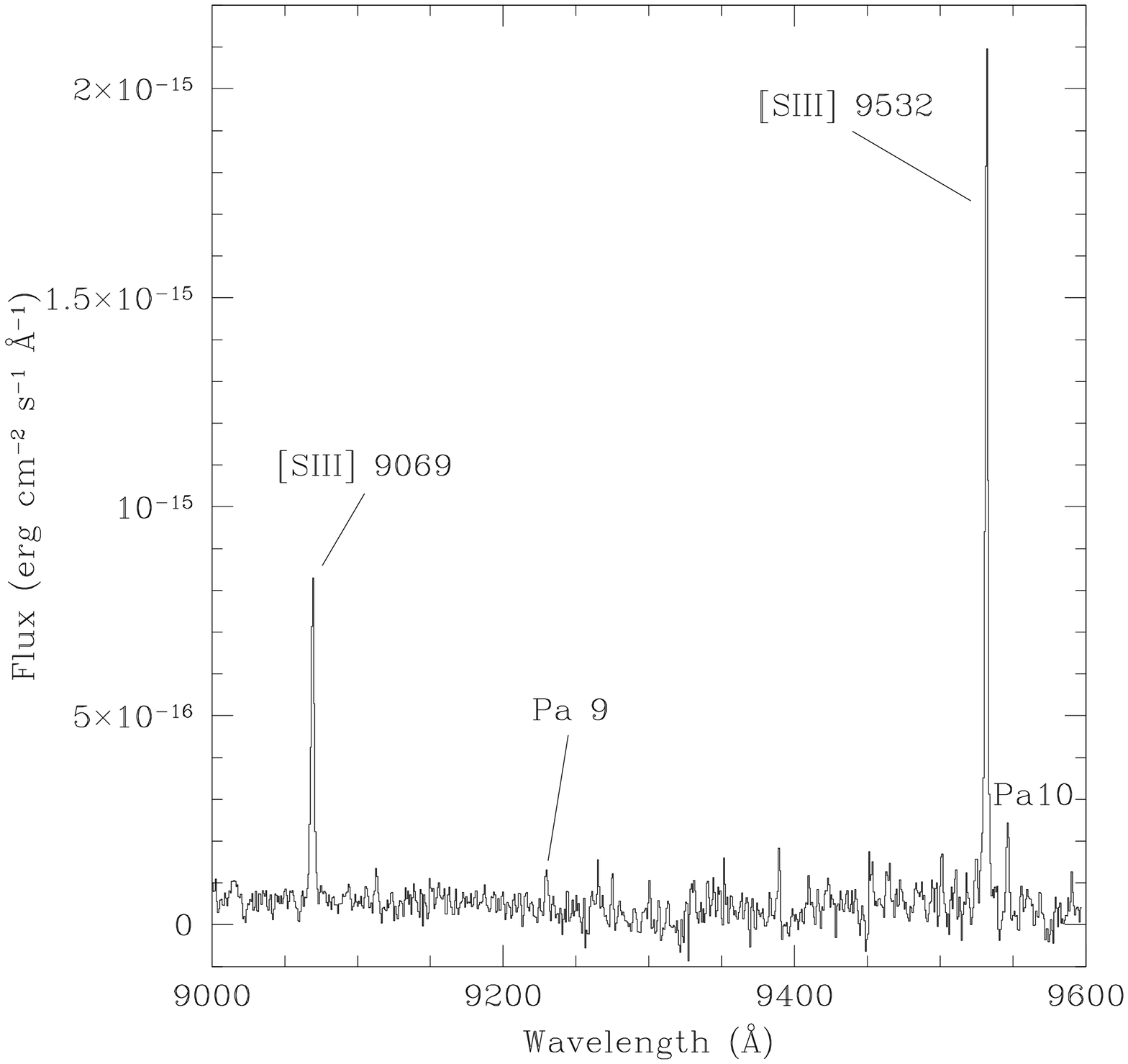}
\includegraphics[width=4.1cm]{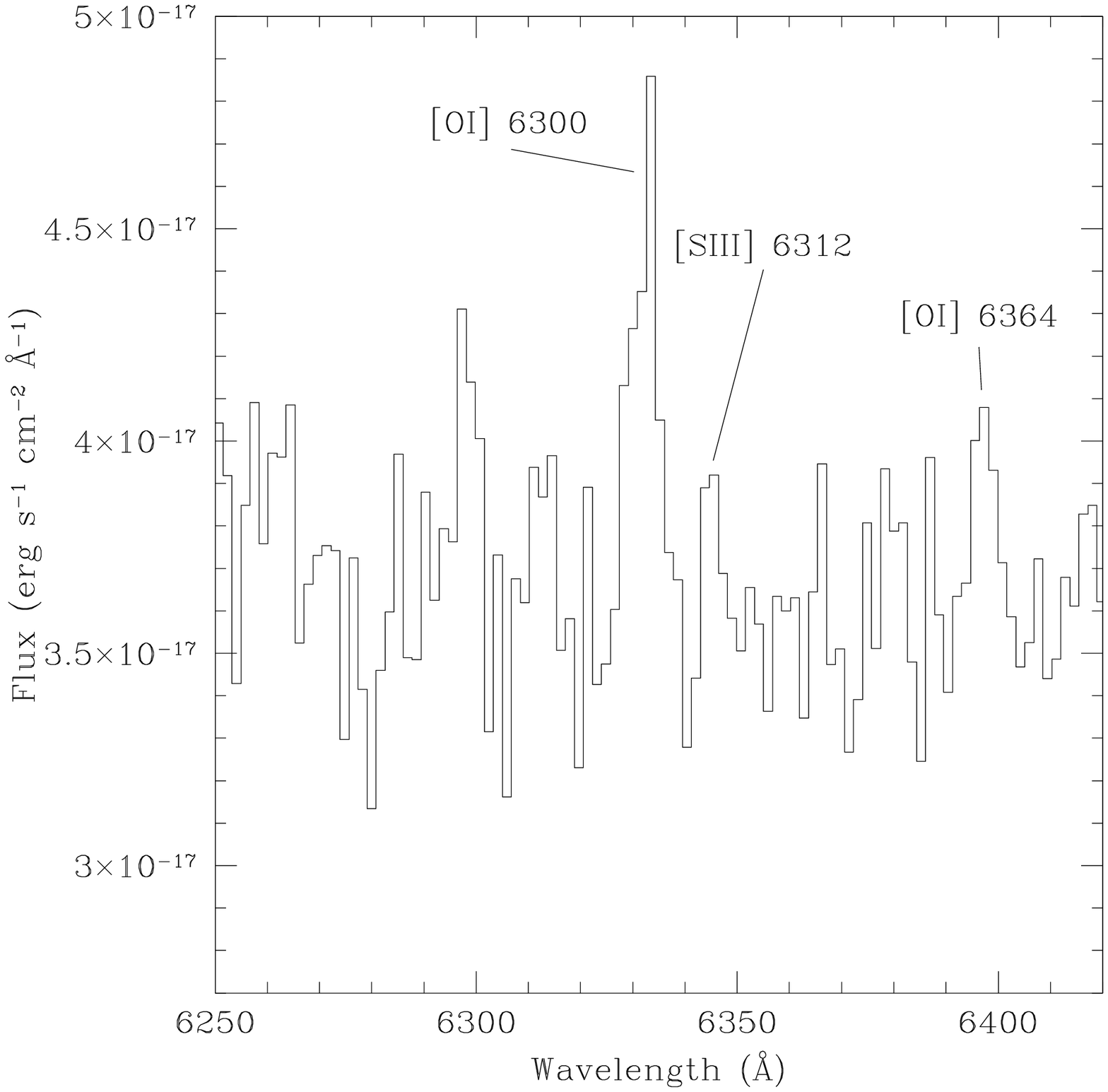}
\end{center}
\figcaption{The spectra of H13, upper pannel, and CDT1, lower pannel, showing the emission lines used for the determination of electron temperatures}
\end{figure*}

On the upper panel, H13 is classified as a GEHR, on the outer disk of NGC 628, a late-type normal spiral galaxy. The derived line temperatures for this region are nearly identical around 10000 K. This diagnostic yields an oxygen abundance of $12+\log(O/H) = 8.24 \pm 0.08$ for the gas (0.2 solar metallicity, if $12+\log(O/H)_\odot = 8.92$). CDT1, in the lower panel, is another GEHR in the inner part of the Sc spiral galaxy NGC 1232, and it seems to be one of the most metallic HII regions where a line temperature has been directly measured. In fact, the derived values for $\log O_{23}$ and $\log S_{23}$ are 0.25 and 0.11 which, according to D\'\i az \& P\'erez-Montero (2000) implies solar or oversolar abundances. The derived values for the $t(S^{2+})$ and  $t(N^+)$ ion-weighted temperatures in this region are 5400 and 6700 K. Hence the found value for the oxygen temperature is $12+\log(O/H) = 8.95 \pm 0.20$ (1.1 solar metallicity)

The derived oxygen abundances for both regions explain the observed differences in the spectra. At high metallicity, the forbidden oxygen lines are far less intense than at low metallicity (as the gas cools down, the oxygen lines dominate in the mid-IR). It is clearly seen that this is not the case for the near-IR sulphur lines due to their lower dependence on the line temperature. Hence, the observation of the near-IR sulphur lines seem to be expendable at high metallicities.

In the case of low excitation regions, it is necessary to rely on different grids of photoionization models in order to infer the value of the non-measured line temperatures. For this purpose, the relations deduced by P\'erez-Montero \& D\'\i az (2002, in preparation) have been adopted. In these relations the new sets of atomic coefficients for $S^{2+}$ (Tayal \& Gupta, 1999) and both the latest versions of stellar atmosphere and photoionization models are taken into account. Hence, the derived relations are compared with those objects for which more than one line temperature has been measured. The relation between $t(O^{2+})$ and $t(S^{2+})$ holds as follows:
\begin{center}
$t(S^{2+}) = (0.90 \pm 0.03)t(O^{2+}) + (0.06 \pm 0.02)$
\end{center}

This approach works out rather well at the low excitation regime, where it could be specially useful, because though the [OIII] $\lambda 4363 \AA$ line cannot be observed, the [SIII] $\lambda 6312 \AA$ can be detected with the aid of both good signal to noise ratio and low-intermediate resolution spectra. However, the validity of this trend at the high excitation regime ($T_e >$ 13000 K) needs to be discussed in further detail. Regarding $t(O^{+}$, $t(N^{+}$ and $t(S^{+}$, the former ones can be taken to be equal to a good aproximation, while the relation with $t(S^{+}$ can be written as follows:

\begin{center}
$t(S^{+}) = (0.85 \pm 0.01)t(O^{+}) + (0.07 \pm 0.01)$
\end{center}

In summary, we have shown that the observation of the near-IR sulphur lines ([SIII], $9069 \AA, 9532 \AA$) can be attained in both low and high excitation HII regions. A moderate spectral resolution (at least $2 \AA/pix$) is required in order to subtract correctly the OH rotational-vibrational transitions of the night-sky spectrum.

\end{document}